\documentclass[aps, twocolumn, superscriptaddress,showpacs, pre]{revtex4-1}
\usepackage{graphicx,amsfonts,amsmath,color,amsbsy,amssymb, subfigure}
\usepackage{epstopdf}

\newcommand{\eps}{\boldsymbol{\epsilon}}
\newcommand{\ets}{\boldsymbol{\eta}}

\begin{document}
\title{Membrane stress and torque induced  by  Frank's nematic textures: A geometric perspective using surface-based
constraints.}

\author{J.A. Santiago}\email{jsantiago@correo.cua.uam.mx}
\affiliation{Departamento de Matem\'aticas Aplicadas y Sistemas\\ 
Universidad Aut\'onoma Metropolitana Cuajimalpa\\
Vasco de Quiroga 4871, 05348 Ciudad de M\'exico, MEXICO}
\affiliation{Departamento de Qu\'imica F\'isica\\ 
Universidad Complutense de Madrid\\
Av. Complutense s/n, 28040, Madrid, SPAIN}
\author{G.  Chac\'on-Acosta}\email{gchacon@correo.cua.uam.mx}
\affiliation{Departamento de Matem\'aticas Aplicadas y Sistemas\\ 
Universidad Aut\'onoma Metropolitana Cuajimalpa\\
Vasco de Quiroga 4871, 05348 Ciudad de M\'exico, MEXICO}
\author{F. Monroy}\email{monroy@quim.ucm.es}
\affiliation{Departamento de Qu\'imica F\'isica\\ 
Universidad Complutense de Madrid\\
Av. Complutense s/n, 28040, Madrid, SPAIN}
\affiliation{Institute for Biomedical Research Hospital Doce de Octubre (imas12)\\
Av. Andaluc\'ia s/n 28041, Madrid, SPAIN}

\vspace{10pt}


\begin{abstract}
An elastic membrane with embedded nematic molecules is considered as a model of anisotropic fluid membrane 
with internal ordering. By considering the geometric coupling between director field and membrane curvature, the 
nematic texture is shown to induce anisotropic stresses additional to Canham-Helfrich elasticity. 
Building upon differential geometry, analytical expressions are found for 
the membrane stress and torque induced by splaying, twisting and bending of the nematic director as described by 
the Frank energy of liquid crystals. The forces induced by prototypical nematic textures are visualized on the sphere 
and on cylindrical surfaces.

\smallskip

\end{abstract}


\maketitle

\section{Introduction}
Nematic textures are intrinsically ordered liquid-crystalline structures  expected to induce non-trivial stresses on 
flexible membranes \cite{hel-prost, kamien, giomi,  santiago}. Their most salient feature is geometric coupling 
between the nematic order and mechanical stress, which  configures
the equilibrium distribution of membrane forces.  Whereas the curvature elasticity of fluid membranes has been 
classically approached from the  Canham-Helfrich (CH) theory \cite{canham, helfrich, yang1, yang2, yang3}, the 
nematic texture can be modeled by the Frank's energy considering the distortion modes of
splaying,  twisting and bending of the nematic director  \cite{gennes}. 
There is a considerable amount of published work on the structural features of liquid-crystalline membranes
both in the theoretical side \cite{napoli, santangelo, kardar, nguyen, aharoni} and in experimental setting 
\cite{New2, New3, New4}, including numerical simulations \cite{New4}.
The present work adds up mechanical details that remain unexplored from a theoretical point of view, especially regarding extrinsic couplings. By adopting a pure geometric standpoint, we contribute an analytic theory for the anisotropic forces induced by the Frank's energy of nematic membranes, which outgoes far beyond the  well-known geometrical theory of fluid membranes \cite{stress, capo, fournier}. Using the new framework for nematic membranes, the emergence of topological forces between defects  could be further analyzed beyond classical approaches \cite{nelson, shin, santiago, lopez-leon, bates}.  The geometric couplings pointed out configure a counterbalance between membrane elasticity and underlying nematic ordering, which gives rise to the distributions of membrane stresses  in dependence with the relative contribution of each material interaction \cite{napoli2010, vergori, napoli2016, napoli2018}.

In our approach, the connections between membrane curvature and nematic ordering are introduced as geometric constraints to equilibrium forces. Technically, geometric coupling is implemented by exploiting the method of auxiliary variables previously developed in the general context of quadratic constraints to membrane geometry \cite{auxiliary, jemal-book}. Specifically, geometrical and compositional constraints accounted for here as Lagrange multipliers contributing to minimize the membrane energy. The curvature-congruent field of membrane stresses is obtained in terms of the Frank's constants for nematic splay ($\kappa_1$), 
twist ($\kappa_2$) and bend ($\kappa_3$), which are defined about the global bending rigidity of the membrane ($\kappa$). The description of the resulting curvature-ordering stresses, hereinafter referred to as the Frank-Canham-Helfrich (FCH) field, should enable not only to detail the distribution of membrane forces but also to obtain evolution equations in membrane systems with intrinsic nematic ordering. The geometric interactions here explicited should become in competition with nematic forces, thus determining the particular shape of the 
flexible membrane,  as previously suggested \cite{nguyen, chen-kamien, mac, jiang, seifert}. We will focus on the  effects imposed by the different Frank's components on membrane stress and torque, which will be derived for typical nematic textures in the spherical and cylindrical curvature settings. To the best of our knowledge, the geometric theory here approached represent a novelty in the physical description of the mechanics of nematic membranes.

The paper is organized as follows: Section \ref{GOE} briefly describes the fundamentals of differential geometry of surfaces that we will be need to later establish our theoretical framework. In Section \ref{STRESS-TENSOR}  the Frank's energy is presented in terms of the surface director field together with the geometrical constraints imposed on it, which determine the  couplings that frame 
nematic membrane energetics. The specific expressions for stress and torque are presented in Section \ref{NEMATICSTRESS}, after detailed calculations in Appendix \ref{APX} using auxiliary variables. In Section \ref{TOTAL}, we introduce the total elastic-nematic stress tensor and the total torque tensor that configurate the core of the mechanical FCH-theory of nematic membranes. In order to visualize the general results in particular cases, we obtain the stress and torque induced by some nematic textures in typical geometric models; in Section \ref{STRESS-SPHERE} for the sphere, and in Section \ref{STRESS-ON-CYLINDER} for the cylinder. In Section \ref{DISS}, the main results are discussed  in the context of the state of the art. Finally,  Section \ref{CON} summarizes  the main conclusions.

\section{Geometry of surfaces}\label{GOE}
\begin{figure}[th]
\centering 
\includegraphics[width=3.2in]{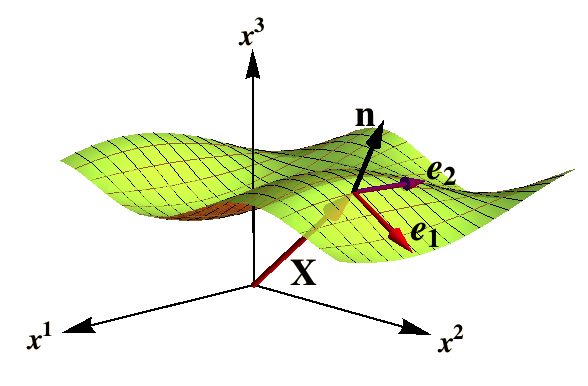}
\vspace{-2mm}
\caption{ The surface ${\bf x}:=(x^1, x^2, x^3 )={\bf X}(\xi^a)$ embedded into $R^3$, the tangent vector fields ${\bf e}_a=\partial_a {\bf X}$,
and the unit normal to the surface ${\bf n}={\bf e}_1\times {\bf e}_2/ \sqrt{g}$; notice that the tangent vector fields may not be
orthonormal  but ${\bf e}_a\cdot {\bf e}_b= g_{ab}$.   
}
\vspace{-2mm}
\label{GASS}
\end{figure}

Let us consider the membrane represented by a differentiable surface manifold embedded into the Euclidean space $R^3$; this surface is defined by the embedding functions ${\bf X}$, which is parametrized by two internal coordinates $\xi^a$, $a=\{1, 2\}$ as
\begin{equation}
{\bf x}= {\bf X}(\xi^a), 
\end{equation}
where the bold denotes the position vector in Cartesian coordinates  ${\bf x}=(x^1, x^2, x^3)$ (see Fig. \ref{GASS}).
 
A local surface basis  can be defined as two vector fields tangent to the surface
${\bf e}_a=\partial_a {\bf X}$,  which define the induced metric 
$g_{ab}={\bf e}_a\cdot {\bf e}_b $. In addition,  $g_{ab}$, and its inverse $g^{ab}$,  respectively 
rises and lowers the tangential indices of surface tensors.

{\it Surface distances}. Figure \ref{GASS} shows the current Riemannian manifold as a metric space where the metric tensor represents the differential distance function; in particular, the length of any tangent vector ${\bf A}= A^a {\bf e}_a $ is given as $ {\bf A}\cdot {\bf A}= |A|^2 =g_{ab} A^a A^b$,    
and the angle $\theta$ between two surface vectors ${\bf A}$ and ${\bf B}$  is determined as 
$\cos\theta = g_{ab} A^a B^b/ |A||B|$. Because a metric is thus available, any derivative can be directly tied to 
the shape of the manifold \cite{spivak}. 
The Christoffel symbols $\Gamma^c_{ab}=\partial_a{\bf e}_b\cdot {\bf e}^c$  provide a  representation 
of the Riemannian  connection in terms of surface coordinates. 
In words, the Christoffel symbols track how the basis changes from
point to point;  they specify intrinsic derivatives along the tangent vectors of the manifold. 
Interestingly, the curve connecting two points that has the smallest length is called a geodesic, 
which fulfills the  equation $\ddot\xi^c  + \Gamma_{ab}^c \dot\xi^a \dot\xi^b=0$ \cite{spivak}. 
Other intrinsic concepts, such as intrinsic curvature, parallel transport, etc., 
can  be expressed in terms of Christoffel symbols. In general, the covariant derivative is refereed to as  
$\nabla_a $,  in terms of surface coordinates $\xi^a$.  In addition, the
unit normal vector to the surface is  defined as ${\bf n}= {\bf e}_1 \times {\bf e}_2/ \sqrt{g}$ 
(where $g=\det g_{ab}$), which complements the local  basis at any point of the surface.

{\it Surface curvatures}. To complete the geometrical description of the surface, we need to define curvatures on the differentiable manifold.  Similarly to the metric tensor needed to measure surface distances, a curvature tensor
$K_{ab}= - \partial_a\partial_b {\bf X} \cdot {\bf n}$ is assigned to each point in the Riemannian manifold. Such $K_{ab}$ 
measures how much the metric tensor is not locally isometric to that of the Euclidean space where 
the surface is embedded. Consequently, the curvature tensor has to be constructed from second derivatives of the 
embedding ${\bf X}(\xi^a)$;  this is $K_{ab}=-\partial_a {\bf e}_b\cdot {\bf n}$. 
At a given point, second order derivatives are connected with the local curvatures through the Gauss equation:  
\begin{eqnarray}
\partial_a {\bf e}_b = -K_{ab}{\bf n}+ \Gamma_{ab}^c{\bf e}_c, \label{GG}
\end{eqnarray}
which involves the  extrinsic curvature $K_{ab}$, 
and the Christoffel symbols $\Gamma_{ab}^c$, associated with the covariant
derivative \cite{docarmo}.  
Whereas the normal components in Eq. \eqref{GG} are said to be {\it extrinsic} - as far they cannot be seen by an 
observer that lives in the surface, the tangential components given by Christoffel symbols are 
purely {\it intrinsic} - since they are only sensed by that internal observer. 
Notice that, using the covariant derivative, the Gauss equation \eqref{GG} can be 
rewritten simply as $\nabla_a {\bf e}_b=- K_{ab}{\bf n}$. 

To specify the geometric connection,  let $U^a$ to be  the tangential components of a  surface vector; then,  
the intrinsic curvature is defined 
as the commutator of the covariant derivatives as:
\begin{equation}
(\nabla_a \nabla_b - \nabla_b\nabla_a)U^c= {\cal R}^c{}_{dab}U^d,
\end{equation}
where ${\cal R}_{abcd}=R_G(g_{ac}g_{bd} - g_{ad}g_{bc})$ is the Riemann tensor and 
${\cal R}_G$ the Gaussian curvature of the surface \cite{docarmo}.

In addition, integrability conditions relate 
intrinsic and extrinsic curvatures througth the Gauss-Codazzi equation
\begin{equation}
K_{ab}K^b{}_c= KK_{ac} - g_{ac}{\cal R}_G, 
\end{equation}
where $K=g_{ab}K^{ab}$. Finally, the Codazzi-Mainardi equation is given by
\begin{equation}
\nabla_a K^a{}_b= \nabla_b K,
\end{equation}
which establishes the structure of the Gaussian map that defines the surface.

{\it Surface derivatives}. With the covariant derivative  $\nabla_a$ and the Gauss equation \eqref{GG}, 
we  obtain the covariant derivative of any surface vector field ${\bf A}$, it can be projected  in the
local basis $\{{\bf e}_a, {\bf n}\}$ as:
\begin{equation}
{\bf A} = A^a {\bf e}_a+ A_n {\bf n},
\end{equation}
where $A^a= {\bf A}\cdot {\bf e}^a$ and $A_n=  {\bf A}\cdot {\bf n}$.
Thus the covariant derivative along the surface can  be expressed   as:  
\begin{equation}
\nabla_a {\bf A}= ( \nabla_a A^b + A_n K_a{}^b ){\bf e}_b + (\nabla_a A_n - A^b K_{ab} ) {\bf n}.\label{CDD}
\end{equation}
Noticeably,  even the tangential component contains the extrinsic curvature of the~surface.

The surface gradient  operator is defined as $\nabla = {\bf e}^a \nabla_a$; when operating  on a scalar function 
defined on the surface, we have
$\nabla f={\bf e}^a\nabla_a f ={\bf e}^a\partial_a f$, which is the surface gradient of the function. 
Using this operator and  Eq. \eqref{CDD}, the surface divergence $\nabla\cdot {\bf A}$ can be written as \cite{aris}:
\begin{eqnarray}
\nabla\cdot {\bf A} &=& {\bf e}^a\cdot \nabla_a {\bf A}, \nonumber\\
&=& \nabla_a A^a + K A_n, \label{DIV}
\end{eqnarray}
which contains the intrinsic divergence $\nabla_a A^a$,  but also an extrinsic term given as $K A_n$. As a matter of fact, using
the Gauss equation, the surface divergence of the unit normal is:
$\nabla\cdot {\bf n}= {\bf e}^a \cdot \nabla_a {\bf n}= K.$

Likewise, the surface curl operator, defined as $\nabla\times={\bf e}^a\nabla_a\times $, can 
be used to obtain
\begin{eqnarray}
\nabla\times {\bf A} &=& {\bf e}^a\times \nabla_a{\bf A}, \nonumber\\ 
&=&( \nabla_a A^b + A_n K_a{}^b ){\bf e}^a\times {\bf e}_b \nonumber\\
&+& (\nabla_a A_n - A^b K_{ab} ){\bf e}^a\times {\bf n}, \nonumber\\
&=& \varepsilon^{ab}\nabla_a A_b \, {\bf n}  +  (\nabla_a A_n - A^b K_{ab} )\varepsilon^c{}_a {\bf e}_c, \label{ROT}
\end{eqnarray}
where we have used Eq. \eqref{CDD} and the antisymmetric  tensor $\varepsilon_{ab}= \sqrt{g}\epsilon_{ab}$
(with $\epsilon_{ab}$ being  Levi-Civita symbols), which defines the 
normal vector ${\bf e}_a\times {\bf e}_b=\varepsilon_{ab}{\bf n}$. Note that  covariant derivatives can be 
substituted by  partial ones. 
As shown by Eq. \eqref{ROT}, the normal component of the curl vector operation is a geometrically intrinsic term.

\begin{figure}[htp!]
\centering 
\includegraphics[width=3.2in]{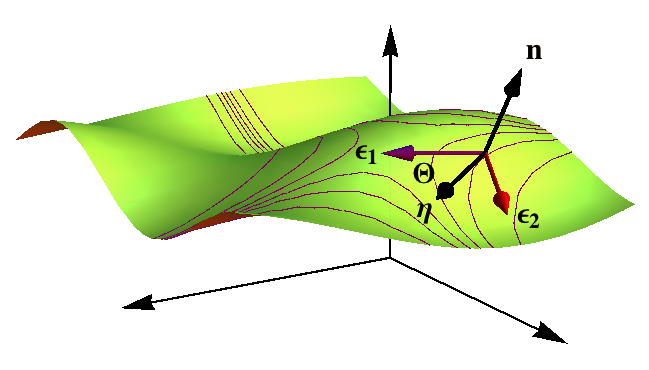}
\vspace{-2mm}
\caption{ Nematic texture on a curved membrane with director  tangent to the surface  such that
$\ets= \eta^\alpha \eps_\alpha$, where $\eps_\alpha$ are two unit tangent vector fields such 
that the unit normal ${\bf n}=\eps_1\times \eps_2$. The vector field $\ets_\perp$ is defined
as $\ets_\perp= \ets\times {\bf n}.$
}
\vspace{-2mm}
\label{VIEL}
\end{figure}

{\it Surface nematic director}. As shown in Figure \ref{VIEL}, 
we can also define  orthonormal vector fields $\eps_\mu$, $(\mu=1, 2)$, tangent
to the surface with $\eps_\mu \cdot \eps_\nu=\delta_{\mu\nu}$ and ${\bf n}={\eps}_1\times \eps_2$. 
Given a surface  vector field $\ets$ we can write it as $\ets= \eta^a {\bf e}_a$, or equivalently as
$\ets= \eta^\mu \eps_\mu$. The director field of the nematic texture  is parametrized as a unit vector field: 
\begin{equation}
\ets = \cos\Theta \eps_1 + \sin\Theta \eps_2,\label{ETTA}
\end{equation}
where $\Theta$ defines its orientation (see Figure \ref{VIEL}) \cite{david}.  Thus,
$\eta^a=\cos\Theta\,  \epsilon_1^a + \sin\Theta\,  \epsilon_2^a $, for the components,  where
we used  the coefficients $\epsilon_\mu^a$  that appear into  the relationship between the basis; these are
$\eps_\mu =\epsilon_\mu^a {\bf e}_a.$

\section{Surface Frank energy  AND GEOMETRIC CONSTRAINTS}\label{STRESS-TENSOR}
For a given texture decorating the surface, the nematic distorsion
energy  is given by  the free Frank's energy \cite{chaikin},
\begin{eqnarray}
H_{\rm Frank} &=& \frac{\kappa_1}{2} \int dA \,   ( \nabla\cdot \ets)^2
+ \frac{\kappa_2}{2} \int dA \,  [ \ets\cdot (\nabla \times \ets) ]^2\nonumber\\ 
& +&\frac{\kappa_3}{2} \int dA\,  [  \ets\cdot \nabla )  \ets     ]^2,\label{FFRR}
\end{eqnarray}
where   the splay,  twist and bend terms are proportional to the respective rigidities ($\kappa_1$, $\kappa_2$ and $\kappa_3$).
These components need to be made explicit in terms of the surface derivatives above described. 
Next, we will  discuss separately the meaning of each component.
   
{\it Splay}. The splay energy density  involves the surface divergence of the nematic director $\nabla \cdot \ets$,
which introduces an energy penalty upon losses of parallel alignment between the elongated molecules \cite{chaikin}.
Assuming that the surface director has no normal component, i.e. $\eta_n = 0$, using
Eq. \eqref{DIV} we can write
$\nabla \cdot \ets = \nabla_a\eta^a.$
Thus, the surface energy due to the splay mode of the nematics is purely intrinsic, i.e. it does not depend 
on how the  surface is embedded in the Euclidean space. 

{\it Twist}.
The twist energy involves  the curl operator as describes the energy penalty upon a  shear distortion 
of the nematic alignment.  
Using the result in Eq. \eqref{ROT} for the surface curl and considering that $\eta_n = 0$, we can write  
\begin{eqnarray}
\nabla \times \ets = (\nabla_a \eta^b) \varepsilon^a{}_b {\bf n} - \eta^b K_{ab} \varepsilon^{ca}  {\bf e}_c.
\end{eqnarray}
Therefore,  the term
$\ets\cdot \nabla \times \ets=\eta^b K_{ab}  \eta^a_\perp$ holds, 
where the vector field $\eta^a_\perp=\varepsilon^{ac}\eta_c$ has been defined  as the tangential 
(in-plane) transverse component of the director field (see Fig. \ref{VIEL}). 
Unlike the splay, the twist energy does depend on the extrinsic curvature. 

{\it Bend}.
The bend  energy density contains the vector field
\begin{eqnarray}
(\ets\cdot \nabla) \ets &=&\ets\cdot {\bf e}^a \nabla_a {\eta}\nonumber\\
&=& \eta^a\nabla_a \eta^b {\bf e}_b - K_{ab}\eta^a\eta^b {\bf n}. 
\end{eqnarray} 
The tangential term (intrinsic) measures the deviation of the nematic director with respect to  geodesic curves.
Conversely, the normal term is purely extrinsic.

{\it Molecular director:  Geometric constraints.}
Once the Frank energy has been completely explicited
as the quadratic moduli of the  distortion modes of the molecular director $\ets$  (see Eq. \eqref{FFRR}), two geometric constraints are specifically taken into account: {\it  i)} the nematic field is completely tangent to the surface ($\ets\cdot{\bf n} = 0$) ; 
{\it ii)} the nematic director is unitary ($\ets\cdot\ets = 1$). Consequently, the total nematic energy:
\begin{equation}
H_N= H_{\rm Frank} +  \int dA\,  \lambda \, \ets\cdot {\bf n} 
+ \frac{1}{2} \int dA \beta \, ( \ets\cdot\ets -1 ),
\end{equation}
where $\lambda$  and $\beta$ are Lagrange multipliers that enforce the constraints  $\ets\cdot {\bf n}=0$, 
and $\ets\cdot\ets=1$, respectively. Since both constraints are local in nature, the corresponding Lagrange multipliers 
are indeed scalar fields defined on the surface. This is because they appear under the integral sign, a formal generalization without practical consequence in the variational evaluation of the equilibrium conditions (see Appendix A).
Although this method was originally introduced to examine generalized
quadratic curvature constraints  \cite{auxiliary}, in the current context will be exploited to 
identify the conserved currents associated to the Euclidean invariance of the nematic energy. 

\section{Nematic Stress and Torque}\label{NEMATICSTRESS}
 The nematic stress tensor ${\bf f}^a_{\rm Frank}$,  appears as a consequence of membrane shape deformations, 
${\bf X}\to {\bf X}+ \delta{\bf X}$, so that
\begin{equation}
\delta_{\bf X} H_N= - \int dA \, {\bf f}^a_{\rm Frank}\cdot \nabla_a \delta {\bf X}. 
\end{equation} 
After integration by parts, we obtain the main property of the stress tensor  as the equilibrium condition; namely, its 
covariant conservation, 
$\nabla_a {\bf f}^a_{\rm Frank}=0$.
The explicit  calculation of the stress  tensor has been presented in the 
Appendix \ref{APX}, 
where we have  found the expression for the Lagrange multiplier 
\begin{eqnarray}
-\lambda&=& {\kappa}_1  \nabla_a \eta^a K + {\kappa}_2 ( 2K_\tau  \nabla_a \eta^a_\perp + \eta^a_\perp\nabla_a K_\tau ) 
\nonumber\\
&&+ \kappa_3 [  \nabla_b \eta^b K_\eta + \eta^b \nabla_b \eta^a K_{ad} \eta^d  \nonumber\\
&&+  \eta^b \eta^c ( \nabla_c \eta^d K_{bd} + \nabla_b K_{ cd}\eta^d + K_{cd} \nabla_b \eta^d      )] ,\label{LMB}
\end{eqnarray}
where  $K_\tau= K_{ab}\eta^a_\perp \eta^b$.
The multiplier $\lambda$  enforces the nematic director to be tangent to the surface. In fact, $\lambda $ contributes to 
the normal force per unit length  on the membrane, as we will see below.
Likewise, the multiplier $\beta$ does not play role in the stress tensor  (see Appendix~\ref{APX}). 

Therefore, the splay stress tensor is found to be:
\begin{eqnarray}
\frac{{\bf f}^a_S}{\kappa_1}&=& \nabla_d \eta^d \left( \nabla^c \eta^a  -
 \frac{g^{ac}}{2}\nabla_d\eta^d \right) {\bf e}_c\nonumber\\
&-&\nabla_c \eta^c (K    \eta^a -  K^a{}_c \eta^c) {\bf n}, \label{UNN}
\end{eqnarray}
where the tangential components contain intrinsic information of the surface through  $g_{ab}$ and 
covariant derivatives of $\eta^a$; the normal component includes, instead, coupling with  extrinsic curvature.
Note that both components are proportional to the divergence of the nematic director, so that in the case of
textures without sources and sinks, the splay stress vanishes.

The force per unit length, on a surface curve with unit tangent ${\bf T}=T^a{\bf e}_a$, and conormal 
${\bf l}= l^a{\bf e}_a$ (see Fig.\ref{ALL}), is calculated by projecting  the stress tensor as \cite{deserno}:
\begin{equation}
{\bf f}^a_S l_a= F_T^S {\bf T}+ F_l^S {\bf l}+ F_n^S {\bf n}, 
\end{equation}
where the projections are given by
\begin{eqnarray}
F_T^S&=& \kappa_1 \nabla_d\eta^d T^c l^a \nabla_c \eta_a, \nonumber\\
F_l^S&=& \kappa_1 \nabla_d\eta^d \left( l^al^c  \nabla_c\eta_a - \frac{1}{2} \nabla_b\eta^b \right),\nonumber\\
F_n^S&=& -\kappa_1\nabla_d \eta^d (  K \eta_a l^a - K_{ab}l^a\eta^b ).
\end{eqnarray}
When considering the twist term of the Frank's energy, using the same method of auxiliary variables (see Appendix \ref{APX}),
the calculated twist stress tensor  is
\begin{eqnarray}
\frac{{\bf f}_W^a}{\kappa_2} &=& K_\tau (K^{cb} \eta_c \eta_\perp^a 
- \frac{g^{ab}}{2}K_\tau  ){\bf e}_b \nonumber\\
&-& [(2K_\tau \nabla_c\eta^c_\perp + \eta^c_\perp \nabla_c K_\tau  ) \eta^a \nonumber\\
&+& K_\tau \eta_{\perp c} \nabla^a \eta^c ] {\bf n},
\end{eqnarray}
where all the terms have coupling with extrinsic curvature.
In this case,  we found:
\begin{eqnarray}
F_T^W&=& \kappa_2 K_\tau K_{cb}\eta^c T^b \eta_\perp^a l_a, \nonumber\\
F_l^W&=& \kappa_2 K_\tau \left(  K_{cb} \eta^c l^b  \eta_\perp^a l_a - \frac{K_\tau}{2}   \right), \nonumber\\
F_n^W&=&- \kappa_2[(2K_\tau \nabla_c\eta^c_\perp + \eta^c_\perp \nabla_c K_\tau  ) \eta^a l_a \nonumber\\
&+& K_\tau \eta_{\perp c} l^a\nabla_a \eta^c ].
\end{eqnarray}
Finally, the bend stress tensor  obtained is:
\begin{eqnarray}
\frac{{\bf f}_B^a}{\kappa_3} &=&  \Big[K_\eta \eta^c \eta^a K^b{}_c + \eta^a\eta^d  \nabla_d \eta^c\nabla^b \eta_c \nonumber\\
&-& \frac{g^{ab}}{2} (\eta^d\eta^e \nabla_d\eta^c \nabla_e \eta_c + K_\eta^2)  \Big] {\bf e}_b \nonumber\\
& -&  \Big[  \nabla_b \eta^b K_\eta + 3\eta^b \nabla_b \eta^c K_{cd} \eta^d  \nonumber\\
&+&  \eta^b \eta^c\eta^d  \nabla_b K_{ cd}  \Big] \eta^a {\bf n}, \label{TERR}
\end{eqnarray}
with the projections 
\begin{eqnarray}
F_T^B&=& \kappa_3[ K_\eta K_{bc}T^b \eta^c + \eta^d \nabla_d \eta^c T^b\nabla_b \eta_c   ] \eta^a l_a,  \nonumber\\
F_l^B&=& \kappa_3[ K_\eta K_{bc}l^b \eta^c + \eta^d \nabla_d \eta^c l^b\nabla_b \eta_c   ] \eta^a l_a \nonumber\\
&-& \frac{\kappa_3}{2}\left( \eta^d\eta^e \nabla_d\eta^c \nabla_e \eta_c + K_\eta^2 \right), \nonumber\\
F_n^B&=&- \kappa_3 \Big[  \nabla_b \eta^b K_\eta + 3\eta^b \nabla_b \eta^c K_{cd} \eta^d  \nonumber\\
&+&  \eta^b \eta^c\eta^d  \nabla_b K_{ cd}  \Big] \eta^a l_a,\label{BENDD}
\end{eqnarray}

{\it Total nematic force.}
Because the energy is additive, we get  the total stress of the nematic membrane as:
\begin{equation}
{\bf f}^a_{\rm Frank}= {\bf f}^a_S + {\bf f}^a_W+ {\bf f}^a_B.
\end{equation}
The analytic outcome in Eqs. \eqref{UNN}-\eqref{BENDD},  is the most
relevant result of this paper.  As far as we know, this result
had not been presented before; it establishes explicit relationships for the tensor components 
of the membrane stress due to the presence of the nematics.

Once we have described the stress tensor, let's look at the consequences of translations and 
rotations in  the energy. Let us notice that when the equilibrium condition is satisfied, the variation
of the energy can be written as 
\begin{eqnarray}
\delta H_N&=& -\int dA \nabla_a ({\bf f}^a_{\rm Frank}\cdot \delta {\bf X}) 
-\int dA \nabla_a ( \Lambda^{ab} {\bf e}_a\cdot \delta{\bf n} )\nonumber\\
&+& \int dA  \nabla_a ( {\cal H}^a \cdot \delta \ets ),
\end{eqnarray}
where we have defined
\begin{equation}
{\cal H}^a= \kappa_1(\nabla_b\eta^b)  {\bf e}^a  
+\kappa_2   K_\tau (\ets\times {\bf e}^a)  
+\kappa_3    \eta^a ( \eta^b \nabla_b \ets ),
\end{equation}
and $\Lambda^{ab}= -\partial H_{\rm Frank}/ \partial K_{ab}$ (see appendix \ref{APX}).

{\it Translations}. Let us consider first an infinitesimal translation of the surface element 
$\delta{\bf X}= {\bf a}$. Deformation of the tangent
vectors can be found,  $\delta {\bf e}_a= \partial_a \delta {\bf X}=0$, 
thus $\delta {\bf n}=0$, and similarly $\delta\ets=0$. Consequently,
\begin{equation}
\delta H_N= - {\bf a}\cdot \oint_{\cal C} ds\, {\bf f}^a_{\rm Frank} l_a, \label{OP}
\end{equation}
and as we mentioned above, ${\bf f}^a_{\rm Frank}l_a$ is identified as the force, per unit length, 
acting on the loop $\cal C$.
The line integral in the right hand of Eq. \eqref{OP} is the generalized force exerted by a surface element decorated with the nematics; otherwise said, it holds for the contribution to membrane tension arising from the nematic texture. This is how translation symmetry give rise to the membrane stress tensor, with the surface tension being the conserved quantity related to this continuos symmetry of the membrane.

{\it Rotations}. Let us consider now an infinitesimal rotation of the shape membrane, 
$\delta{\bf X}= \boldsymbol{\omega} \times {\bf X}$. The unit normal undergoes a rotation,
$\delta {\bf n}= \boldsymbol{\omega}\times {\bf n}$, and the nematic director 
changes as $\delta \ets=- (\ets\cdot \delta{\bf n})\,  {\bf n}$. Consequently, 
under an infinitesimal rotation, the  energy deformation can be written as 
\begin{eqnarray}
\delta H_N &=& - \boldsymbol{\omega}\cdot \int dA \nabla_a {\bf M}^a_{\rm Frank}, \nonumber\\
&=&  -\boldsymbol{\omega}\cdot \oint_{\cal C} ds \, {\bf M}^a_{\rm Frank}l_a,
\end{eqnarray}
where the nematic torque is defined as
\begin{equation}
{\bf M}^a_{\rm Frank} = {\bf X}\times {\bf f}^a_{\rm Frank} + {\bf m}^a_{\rm Frank}, \label{EF}
\end{equation}
and 
\begin{equation}
{\bf m}^a_{\rm Frank}=(\kappa_2 K_\tau \eta_\perp^a + \kappa_3  K_\eta \eta^a ) \ets_\perp. 
\end{equation}
The first term in Eq. \eqref{EF} is the external nematic torque, induced by the Frank energy;
the second one is identified as the intrinsic nematic  torque, a vector field that  points 
in direction $\ets_\perp$,  the rotational axis.
We notice that the splay energy does not induce intrinsic torque, this is because under rotations, 
the nematic director deforms in a normal direction to the surface.
The nematic torque here obtained is the second foremost  result of this paper.\\
\section{Total elastic-nematic stress and torque}\label{TOTAL}
\begin{figure}[th]
\centering 
\includegraphics[width=2.0in]{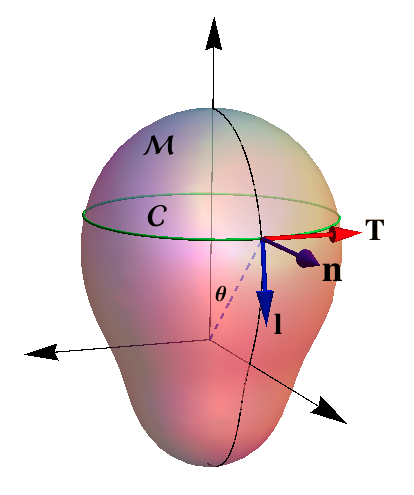}
\vspace{-2mm}
\caption{The Darboux frame adapted to the curve $\cal C$ \cite{docarmo}: 
$\bf T$ the unit tangent, $ \bf n $ the unit 
normal to the surface and the conormal
${\bf l}= {\bf T} \times {\bf n}$. We  can be expand along the tangent basis to the surface as 
${\bf T}=T^a{\bf e}_a$, and   ${\bf l}=l^a{\bf e}_a$, so that $T^a T_a=1=l^al_a$ and $ T^al_a=0$. The induced 
metric can then be written as $g_{ab}= T_aT_b + l_al_b$.}
\vspace{-2mm}
\label{ALL}
\end{figure}

After calculating the nematic components to the stress tensor, contributions from surface elasticity must be 
properly casted to account  for the total membrane stress; the complete free energy  is given by
\begin{eqnarray}
H&=& H_N +  H_{\rm CH}, \label{FTT}
\end{eqnarray}
where the Canham-Helfrich (CH) energy holds
\begin{equation}
H_{\rm CH}= \frac{\kappa}{2}\int dA (K-K_0)^2 + \sigma \int dA.\label{CANHEL}
\end{equation}
This functional accounts for the flexural elasticity and the lateral tension of fluid membranes, 
which can be described  in terms of surface geometry through the bending rigidity $\kappa$, the spontaneous curvature 
$K_0$, and the membrane tension $\sigma$ \cite{seifert}. 
After minimization $\delta F_{\rm CH}=0$, the CH stress tensor,  can be written as~\cite{deserno}:
\begin{eqnarray}
{\bf f}^a_{\rm CH}&=&\Big\{ \kappa (K-K_0)[ K^{ab}-\frac{1}{2}(K-K_0)g^{ab}  ] 
-\sigma g^{ab} \Big\} {\bf e}_b \nonumber\\
&-& \kappa( \nabla^a K) {\bf n}. 
\end{eqnarray}
By taking this result   into  account, the total stress tensor of the nematic membrane energetically described by  Eq. \eqref{FTT} 
is thus ${\bf f}^a = {\bf f}^a_{\rm Frank} + {\bf f}^a_{\rm CH}$.
For a  closed membrane, namely, a vesicle,  the 
pressure difference $P$ between the outer medium and the vesicle interior imposes that the stress tensor 
is not conserved but 
\begin{equation}
\nabla_a {\bf f}^a= P{\bf n}.\label{FGH}
\end{equation}
When integrating Eq. \eqref{FGH} over the area of the patch ${\cal M}$ with boundary the 
loop ${\cal C}$, parametrized by arc length $s$,  as showed in Figure \ref{ALL}, 
we have
\begin{eqnarray}
\oint_{\cal C} ds\, {\bf f}^al_a &=&P \int_{\cal M}\,  dA\,  {\bf n},\nonumber\\
&=&  \frac{P}{2} \oint_{\cal C}\,  ds \,  {\bf X}\times {\bf T}.
\label{EQQ}
\end{eqnarray}
Let's notice that this equation sets down a counterbalance between the compositional stress due to the nematics and the
elastic  stress due to bending stiffness, membrane tension and external pressure.  
Because the force that the nematic contained 
in region $\cal M$ exerts on the loop $\cal C$    
is given by
\begin{equation}
{\bf F}_{\rm Frank}= \oint_{\cal C} ds\,   {\bf f}^a_{\rm Frank} l_a,  
\end{equation}
where  the nematic force, per unit length, is
\begin{eqnarray}
{\bf f}^a_{\rm Frank}l_a&=&({\bf f}^a_S + {\bf f}^a_W  +{\bf f}^a_B ) l_a,\nonumber\\
&=& F_T^{\rm Frank} {\bf T} + F_l^{\rm Frank} {\bf l} + F_n^{\rm Frank} {\bf n},
\end{eqnarray}
and  $ F_T^{\rm Frank}= F_T^S + F_T^W + F_T^B$; similarly for $F_l^{\rm Frank}$ and~$F_n^{\rm Frank}$.\\
In order to account for the equilibrium tradeoff between ordering interactions and membrane elasticity, the 
nematic force have to be completed  with the elastic force given by \cite{fournier, deserno}
\begin{eqnarray}
F_n^{\rm CH}&=&-\kappa \nabla_l K, \nonumber\\
F_l^{\rm CH}&=& -\Sigma +\frac{\kappa}{2} [K_l^2 - K_T^2 + 2K_T K_0 ],\nonumber\\
F_T^{\rm CH }&=& \kappa ( K-K_0 )K_{lT}, \label{DF}
\end{eqnarray}
where the effective membrane tension $\Sigma= \sigma + \kappa K_0^2/2$, $ K_l= K_{ab}l^al^b$, $K_T= K_{ab}T^aT^b$,
$K_{lT}=K_{ab}l^a T^b$ and $\nabla_l K=\l^a\nabla_a K$. 

Furthermore, the total  torque can be written as
\begin{eqnarray}
{\bf M}^a &=& {\bf M}^a_{\rm CH} + {\bf M}^a_{\rm Frank},\nonumber\\
&=& {\bf X} \times {\bf f}^a + {\bf m}^a,
\end{eqnarray}
where the total intrinsic torque is 
\begin{equation}
{\bf m}^a={\bf m}^a_{\rm CH} + {\bf m}^a_{\rm Frank},
\end{equation}
and ${\bf m}^a_{\rm CH}=- \kappa ( K-K_0) {\bf T}$, the intrinsic torque induced by the
CH energy \cite{deserno}.

Behind the completeness of these results,  become straightforwardly simplified in highly symmetric  
geometric settings, e.g. the sphere or the cylinder. In case
the director field lines up along a principal direction of the surface, 
namely $\kappa_p $, i.e. $K^a{}_b \eta^b_p= \kappa_p \eta^a_p$, thus
$K_\eta= \kappa_p$ and  $K_\tau=0$; consequently, the twist force vanishes. Below, 
we examine the surface distribution of total stresses in the particular cases of 
the sphere (Section \ref{STRESS-SPHERE}), and the cylinder geometry (Section \ref{STRESS-ON-CYLINDER}). 

\section{Stress on spherical vesicles}\label{STRESS-SPHERE} 
An interesting geometry relevant to the mechanics of minimal cells \cite{boal, monroy}, 
is a spherical vesicle coated with a nematic
texture \cite{vinel}. In spherical coordinates the induced metric determines the line element as
$ g_{ab}d\xi^a d\xi^b= R^2 d\theta^2 + R^2\sin^2\theta\, d\phi^2 $. Let's consider
the loop $\cal C$ to be  the spherical parallel with polar angle $\theta$, 
$\cal M$  being the patch up to  $\theta_0$ on the north hemisphere
as depicted in Figure \ref{ALL}. For this curve we have 
${\bf T}=\boldsymbol{\phi}$, and ${\bf l}=\boldsymbol{\theta}$, so that
$T_\theta=0=T^\theta$ and $T_\phi=R\sin\theta=(T^\phi)^{-1}$, 
$l_\theta=R=(l^\theta)^{-1}$ and $l_\phi=0=l^\phi$. For this path, we have $X_T=0$,   
$X_n= R $ and $X_l= 0$,
and thus the local  balance  in Eq. \eqref{EQQ},  is determined as
\begin{eqnarray}
F_T&=&0, \nonumber\\
F_n&=& 0 ,\nonumber\\
F_l&=& - \frac{PR}{2}.\label{LK}
\end{eqnarray}
whether the director  field does not depend on the azimuthal angle $\phi$ (revolution symmetry), 
the local equilibrium condition eq.\eqref{EQQ} gets into
\begin{equation}
 -F_l  + F_n \cot\theta   =  \frac{PR}{2},  \label{SSRT}
\end{equation}
where the functions $F_l$ and $F_n$ both depend on the nematic texture.
\subsection{Nematic texture with $\Theta=\pi/2$}
\begin{figure}[th]
\centering 
\includegraphics[width=3in]{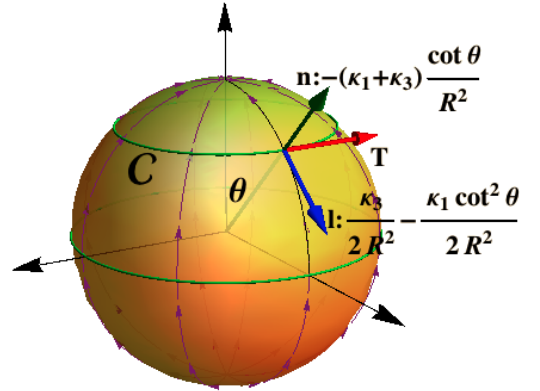}
\vspace{-2mm}
\caption{Nematic texture with $\Theta=\pi/2$, and the induced local forces $F_l$ and $F_n$; with this texture, 
the equilibrium equation is given by Eq. \eqref{YYLS}.} \label{LCL} 
\label{FF2}
\end{figure}

This particular case represents a nematic director oriented along the spherical meridians (see Fig.\ref{LCL}).
Because the director field can be  written in terms of the orthonormal basis as in Eq. \eqref{ETTA},
if we take $\eps_1= {\bf T}$ then $\eps_2= -{\bf l}$. Consequently,  the nematic 
texture with $\Theta=\pi/2$ implies that $\ets= \eps_2=-{\bf l }$ and $\ets_\perp= {\bf T}$.
After some algebra, one gets:
\begin{equation}
\nabla_a\eta^a= -\frac{1}{R}\cot\theta, \label{DVV}
\end{equation}
but $T^c l^a \nabla_c\eta_a= \Gamma^\theta_{\phi\theta}/ R\sin\theta=0$,  thus $F_T^S=0$.
For the longitudinal direction, we see that $l^al^c\nabla_a\eta_a=\Gamma^\theta_{\theta\theta}/R=0$, 
thus $F_l^S= - \kappa_1\cot^2\theta /2R^2$. Finally, along the normal we have
$K\eta_a l^a = -2/R$ and $K_{ab}l^a \eta^b=-1/R$ and then
$F_n^S=-\kappa_1\cot\theta/R^2$.
Regarding the bending component along the meridians, $K_\eta=1/R$ and $K_{ab}T^a\eta^b=0$, so we get 
$\eta^a\nabla_a \eta^c=\Gamma^c_{\theta\theta}/R^2=0$. Therefore, we deduce  $F_T^B=0$ and 
$F_l^B= \kappa_3/2R^2$ and, since $\nabla_\theta K_{\theta\theta}=0$, we have
$F_n^B= -\kappa_3 \cot\theta/ R^2$. The twist component of the stress tensor is found to vanishes on the
sphere, i.e. ${\bf f}^a_W=0$, and it does not induce forces at all. 
Finally we found the Darboux components of the total  force as:
\begin{eqnarray}
F_T&=&0,\nonumber \\
F_l&=&-\Sigma + \frac{\kappa K_0}{R} - \frac{ \kappa_1}{ 2 R^2   }  \cot^2 \theta    + \frac{\kappa_3}{2 R^2}, \nonumber\\
F_n&=&- \frac{1}{R^2} \left(\kappa_1+\kappa_3 \right) \cot\theta. \label{PPA}
\end{eqnarray}
According to Eq. \eqref{DF},  no contribution from the bending stiffness is expected in the sphere at zero 
spontaneous curvature ($K_0=0$).
However, looking at the total force $F_l$ in Eq. \eqref{PPA}, a splay component  of magnitude $\kappa_1\cot^2\theta/2R^2$,
must be stressed in order to make a sectional cut. Particularly, its strength is
$\kappa_1/2R^2$ at the equatorial loop, becoming more intense as the cut approaches to the poles.
Furthermore,  a constant force $\kappa_3 /2R^2$  is induced by the bending of the 
nematic director,  but director twisting does not affect anymore, as expected for a spherical  texture that  
circulates along meridians,  avoiding rotation between the poles. Similarly,  the normal force $F_n$, 
does not depend  on $\kappa_2$ anymore; if splay and bending terms are taken into account, 
it vanishes at the equatorial loop ($\theta = \pi/2$),  and diverges as approaching to the poles ($\theta = 0, \pi $). This normal force  $F_n$ is radial and directed towards the interior on the northern hemisphere, while directed outward 
on the southern, which causes a dipolar imbalance between the two hemispheres. After substituting Eq. \eqref{PPA}
in Eq. \eqref{SSRT},  we get
\begin{equation}
\frac{PR}{2}=\Sigma -\frac{\kappa K_0}{R}- \frac{\kappa_1}{2R^2}\cot^2\theta  - \frac{\kappa_3}{2R^2}( 1+ 2\cot^2\theta),\label{YYLS}
\end{equation}
which establishes the equilibrium condition.
An alternative way to rise this condition consists to analyze   the forces separately.  
Under the integral definition in Eq. \eqref{EQQ}, by taking the leftmost hand side for the total
force we found
\begin{equation}
{\bf F}(\theta)= 2\pi R \sin\theta ( - F_l  \sin\theta
+ F_n \cos\theta  ) {\bf k}. \label{FTOT}
 \end{equation}
 Further, separating this force into components  ${\bf F}(\theta)=  {\bf F}_S + {\bf F}_W +  {\bf F}_B + {\bf F}_E$
where the subindices refer to  splay, twist, bend  and elastic forces, respectively given by 
\begin{eqnarray}
{\bf F}_S&=& -\frac{\pi }{R} \kappa_1 \cos^2\theta {\bf k} , \label{SPLAY}       \\
{\bf F}_W&=& 0,\\
{\bf F}_B&=& - \frac{ \pi  }{  R }\kappa_3 (2-\sin^2\theta  ) {\bf k}. \label{BEND}\\
{\bf F}_E&=& 2\pi R \left( \Sigma - \frac{\kappa K_0}{R} \right)\sin^2\theta\,  {\bf k}.
\end{eqnarray}
Notice that the finite values of the splay force is maximum at the poles and  vanishes on the equatorial loop. As deduced above, the twist force is exactly null in the spherical configuration.  The bending force is in general non zero
but minimal at the equator ($\theta=\pi/2$). Finally, the  bend-force goes downward, 
along the opposite direction to the surface tension; if the loop is close the pole, the bend-force is twice than on the equatorial loop. The force induced by the Laplace pressure  $P$  on the parallel loop ${\cal C}$ in Fig. \ref{LCL},  is given  by
\begin{equation}
{\bf F}_P= - P \pi R^2 \sin^2 \theta\,  {\bf k}, 
\end{equation}
therefore, the equilibrium condition $\sum {\bf F}_i=0$,   implies Eq.\eqref{YYLS}.
\\

For this nematic texture, the  intrinsic torque on parallels is given by
\begin{equation}
{\bf m}^a l_a = -\left[  \frac{\kappa_3}{R} + \kappa \left(  \frac{2}{R}-K_0  \right)      \right]\boldsymbol{\phi}.
\end{equation}
Therefore, the couple due to the nematic texture adds to the bending one, and counteracts
to the effect of the spontaneous curvature. On  the meridians, the intrinsic nematic torque vanishes.

\subsection{Nematic texture with $\Theta=0$}
\begin{figure}[th]
\centering 
\includegraphics[width=2.7in]{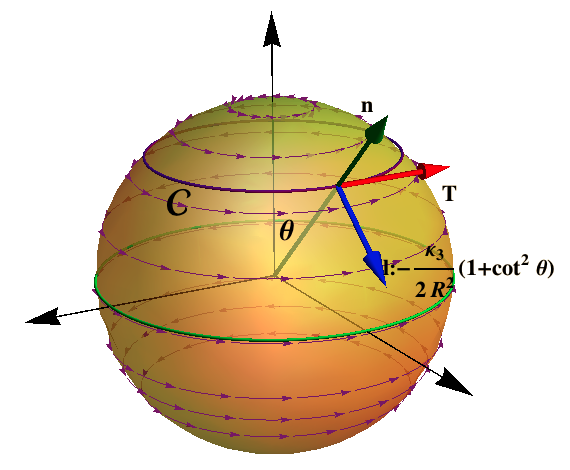}
\vspace{-2mm}
\caption{The nematic texture $\Theta=0$,  on the  sphere. 
The only non trivial local force points  along the conormal $\bf l$.}
\label{SPHE4}
\end{figure}

This texture represents an orientation along spherical parallels; 
here, we identify $\ets = {\bf T} $ and $\ets_\perp= {\bf l}$~(see Fig.  \ref{SPHE4}).
In this case, the divergence of the nematic field vanishes, thus the
nematic texture does not induce splay forces;  let's  notice   that  $\eta^a l_a=0$.
Consequently, the non-trivial terms  are $\eta^a \nabla_a \eta^\theta=- \cot\theta/R^2$, 
and the normal curvature at the parallel, $K_\eta=1/R$, which determines the longitudinal
components of the  bending force, $F_l^B$ in Eq. \eqref{BENDD}. As a consequence, 
the total force points along the conormal ${\bf l}$. Once the elastic force is considered, 
we found
 \begin{equation}
F_l =-\Sigma + \frac{\kappa K_0 }{R} - \frac{\kappa_3}{2 R^2} ( 1+ \cot^2\theta ).\label{FG}
\end{equation}
The bend force on the entire loop (except for them close to the poles) is a constant and points upward as
\begin{equation}
{\bf F}_B= \frac{\pi \kappa_3}{R}  {\bf k}.
\end{equation}
In this case, the equilibrium equation can be expressed as
\begin{equation}
\frac{PR}{2}=\Sigma - \frac{\kappa K_0}{R} + \frac{\kappa_3}{2 R^2}\left( 1+\cot^2\theta   \right),
\end{equation}
which only contains elasticity and nematic bending terms playing against each other. Splay and twist terms are missing in this case as they do not contribute to distort the parallels.

Because on parallel loops, the condition $\ets\cdot {\bf l}=0$ holds, here, the intrinsic nematic torque vanishes and the total intrinsic torque is given by
\begin{equation}
{\bf m}^a l_a = - \kappa\left( \frac{2}{R} -K_0 \right)\boldsymbol{\phi}.
\end{equation}

On meridians, we can find instead
\begin{equation}
{\bf m}^a l_a = -\left[  \frac{\kappa_3}{R} +\kappa \left(  \frac{2}{R}-K_0  \right)      \right]\boldsymbol{\theta}.\label{FK}
\end{equation}
Therefore, the nematic couple adds up to the bending one and counteracts the effect of 
the spontaneous curvature.

\section{Stress on  cylindrical surfaces}\label{STRESS-ON-CYLINDER}
\begin{figure}[th]
\centering 
\includegraphics[width=2.2in]{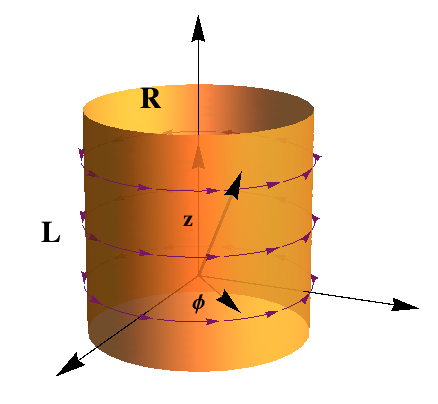}
\vspace{-2mm}
\caption{The nematic texture $\eta^\phi = 1/R$ and $\eta^z=0$ on the cylinder,
if $L$ is fixed then $R=\sqrt{ (\kappa+  \kappa_3)/ 2\Sigma}$.
}
\label{NMY}
\end{figure}

\begin{figure}[th]
\centering 
\includegraphics[width=2.2in]{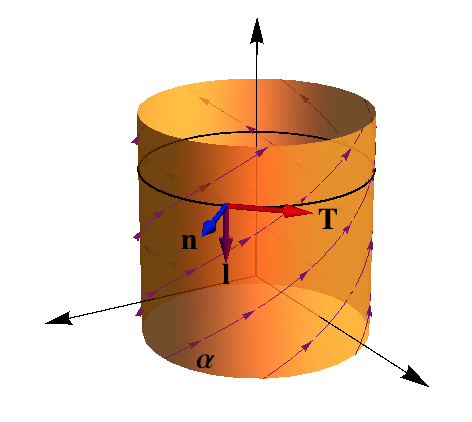}
\vspace{-2mm}
\caption{The nematic texture \eqref{CLL}  on the cylinder,
$\alpha$ being the angle of the nematic director with the horizontal.
The only non trivial local force points  along the conormal ${\bf l}$  in eq.\eqref{FSE}.
On this loop,  $T^\phi=1/R$, $T^z=0$, $l^\phi=0$, $l^z=-1$, are the
components of the Darboux basis.
}
\label{CLYY}
\end{figure}

A cylindrical surface of radio $R$ and length $L$ can be parametrized as  
\begin{equation}
{\bf X}(\phi, z)= (R\cos\phi, R\sin\phi, z ),
\end{equation}
where the azimuthal angle varies in the full domain $0\leq \phi < 2\pi$,  and $z\in[-L/2, L/2 ]$.
The tangent vectors are expressed as ${\bf e}_\phi= R\boldsymbol{\phi} $  and ${\bf e}_z={\bf k}$, whereas the 
unit normal is ${\bf n}=(\cos\phi, \sin\phi, 0)$. On the surface $ds^2= dz^2 + R^2 d\phi^2$, gets the infinitesimal distance whereas
$K_{\phi\phi}= R$ and $K_{zz}=0, K_{z\phi}=0$ are the components of the extrinsic curvature.
In this geometry,  if the nematic texture is 
directionally aligned on meridians, the Frank  energy exactly vanishes. 
Consequently, with the exception of the pure elastic force, no additional force  have to be overcome
to suction a tube in a cylindrical micropipette.
Unlike, if the nematic director aligns with parallels, the components are 
$\eta^\phi = 1/R$ and $\eta^z=0$ (see Fig. \ref{NMY}),  and thus
the bend energy density becomes $\kappa_3/2R^2$. In this case
the membrane energy reads in terms of the tube dimensions as
\begin{equation}
H= \sigma A +\frac{\kappa A}{2} \left( \frac{2\pi L}{A}- K_0  \right)^2+    \frac{\kappa_3}{2}\frac{4\pi^2 L^2}{A},
\end{equation}
where $A= 2\pi R L$ is the cylinder area.
If the tube length is fixed at $L$, then the condition $\partial_A H= 0$ determines the 
equilibrium radius  at $R_{\rm eq}=\sqrt{ (\kappa+  \kappa_3)/ 2\Sigma}$,
where the energy reaches a minimum  and we have introduced $\Sigma= \sigma + \kappa K_0^2/2$. 
In the absence of bending nematics, this formula reduces to the classical result for the 
equilibrium radius of  a lipid tube, 
$R_{\rm eq} = \sqrt{\kappa / \Sigma}$.  \\
A more general result, can be also obtained by cancelling out   the stress around the symmetry axis.
For a general texture as in Fig. \ref{CLYY}, let's consider
\begin{equation} 
\ets=\cos\alpha\, \boldsymbol{\phi} + \sin\alpha {\bf k}, \label{CLL}
\end{equation} 
to be the nematic director   oriented at an angle $\alpha$ with respect to the 
unit azimuthal vector $\boldsymbol{\phi}$ ( see Fig.  \ref{CLYY}).
The components are then $\eta^\phi=\cos\alpha/R$ and $\eta^z=\sin\alpha$. We also 
see that $\ets_\perp= \sin\alpha\, \boldsymbol{\phi} -\cos\alpha\, {\bf k}$.
Additionally, since $\eta_al^a=-\sin\alpha$, $\eta^a_\perp l_a= \cos\alpha$,  
$\eta^bT_b=\cos\alpha $,  $K_{ab}\eta^al^b=0$, $K_{ab}\eta^aT^b=\cos\alpha /R$,
one obtains
\begin{eqnarray}
\nabla_a \eta^a &=& 0, \nonumber\\
K_\eta &=&  \frac{\cos^2\alpha}{R}, \nonumber\\
K_\tau&=& \frac{\sin\alpha\cos\alpha}{R}.
\end{eqnarray}
Because the divergence of the texture is  exactly null in this case, the splay 
force vanishes as well. Therefore, the components of the local  force
on the loop can be written as
\begin{eqnarray}
F_T&=&\frac{\kappa_2}{R^2}\sin\alpha \cos^3\alpha, \nonumber\\
F_l&=& -\frac{\kappa_2}{2R^2}\sin^2\alpha \cos^2\alpha - \frac{\kappa_3}{2R^2}\cos^4\alpha,  \nonumber\\
F_n&=&0. \label{DH}
\end{eqnarray}
In a circular parallel loop, the normal force is found identically zero, as expected for a tube at mechanical equilibrium. However, the two in-plane components adopt non-trivial dependences on the orientation $\alpha$ of the texture. Whereas the longitudinal component is affected by twisting and bending nematic terms, the tangential component that maintains the circulation around the tube axis is exclusively determined by twisting.
The total force on the entire loop 
is given by
\begin{equation}
{\bf F}= 2\pi R\, ( F_l\,  {\bf l}  + F_T {\bf T}).
\end{equation}
In the particular case of the parallel texture,  $\alpha=0$ (see Fig. \ref{NMY})   only the bending force determines the longitudinal 
component $F_l= -\kappa_3/ 2R^2$; in the opposite case, $\alpha= \pi/2$,  then  $F_l=0$. With the 
intermediate texture, 
$\alpha=\pi/4$,  we found $F_T= \kappa_2 /4R^2$ and  $F_l=-\frac{1}{8R^2} ( \kappa_2 + \kappa_3 ).$\\
We further consider the case of meridian directions along the cylinder axis, where ${\bf T}= {\bf k}$,  thus  $T^z=1$ and $T^\phi=0$;
here ${\bf l}= {\bf e}_\phi/R$ and  $l^\phi=1/R$, thus $K_l=1/R$ and $K_T=0$. 
Therefore, $\eta^a l_a=\cos\alpha$, $\eta^b T_b= \sin\alpha$, $ K_{ab}\eta^a l^b= \cos\alpha$, 
$K_{ab}\eta^a T^b=0$. Consequently, the components of the force are found as 
\begin{eqnarray}
F_T&=& 0, \nonumber\\
F_l&=& -\frac{\kappa_2}{2R^2}\sin^2\alpha \cos^2\alpha + \frac{\kappa_3}{2R^2}\cos^4\alpha,\nonumber\\
F_n&=&0, \label{FSE}
\end{eqnarray}
i.e. only the longitudinal component is non zero in this case.
In fact, we can adjust the radius of the cylinder, such that the stress on any meridian vanishes. 
This experiment could consist to fix the nematic texture at rotational orientation ($\alpha=0$),  
then $R_{\rm eq}= \sqrt{(\kappa + \kappa_3) /2\Sigma}$,  independently of the splay and twisting rigidities.
In the uniaxial orientation ($\alpha=\pi/2$), the equilibrium radius reduces becomes exclusively 
determined by membrane elasticity, i.e.
$R_{\rm eq}= \sqrt{\kappa/ 2\sigma}$, which determines a much smaller tensions than necessary to 
realize the rotational texture. In the most general case, we found tubes with equilibrium radius
\begin{equation}
R_{\rm eq}=\sqrt{ (\kappa + \kappa_3 \cos^4\alpha -\kappa_2 \sin^2\alpha \cos^2\alpha)/ 2\Sigma}.
\end{equation}
On parallels, the torque can be written as
\begin{eqnarray}
{\bf m}^al_a &=& \Big[ \frac{\sin\alpha\cos\alpha}{R}( \kappa_2 \cos^2\alpha - \kappa_3 \sin^2\alpha) \nonumber\\
&-& \kappa\left( \frac{1}{R} - K_0  \right) \Big] \boldsymbol{\phi} \nonumber\\
&-&\frac{\cos^2\alpha}{R} ( \kappa_2 \cos^2\alpha - \kappa_3 \sin^2\alpha ) {\bf k}.  
\end{eqnarray}
The nematic texture with $\alpha=\pi/ 2$, does not induce nematic intrinsic torque, as expected.   
If $\alpha=0$,  we find $ {\bf m}^al_a= - \kappa (1/R -K_0 ) \boldsymbol{\phi} - \kappa_2/R\,  {\bf k}$, and the twist
does not play  in the couple. Nevertheless, any deviation of the circular texture,  induces a contribution 
of the twist; if $\alpha=\pi/4$ we have
\begin{equation}
{\bf m}^a l_a  =\frac{1}{4R} ( \kappa_2 - \kappa_3 -4\kappa) \boldsymbol{\phi} - 
\frac{1}{4R} ( \kappa_2 -\kappa_3) {\bf k}.\label{DK}
\end{equation}
Importantly,  the local torque induced by the twist counteracts the one induced by the bending mode of the director field.

\section{discussion}\label{DISS}
We have obtained analytical expressions for the stress tensor and the torque
of an elastic membrane decorated with a nematic texture constrained to lie tangent to the membrane. 
The nematic texture is modeled
by the Frank energy, which takes into account splay,  twist and bend orientations of the nematic director.
A pure geometric standpoint is adopted to get the distribution of internal forces due to local coupling between the nematic field and membrane curvature. Specifically, the geometric characteristics of the nematic field (tangential $\ets\cdot {\bf n} = 0$, and normalized $\ets\cdot\ets = 1$), together with the surface metric and its curvature, are connected to the embedding function that defines the surface by introducing auxiliary variables that impose the  appropriate constraints. The method of auxiliary variables, previously developed by Guven to impose geometric congruence for generalized quadratic 
constraints \cite{auxiliary, jemal-book}, has been here re-adapted to identify the components of the surface 
stress tensor after the different terms of the nematic field. 
The structure of the membrane stress tensor exhibits a non-trivial interplay between  nematic structure
and geometry, showing  the relevance of the extrinsic effects, particularly of the coupling
between  nematic director and  extrinsic membrane curvature. For instance, whereas
the surface stress induced by nematic splaying  exhibits a  tangential component with 
only intrinsic coupling, the normal component is chiefly driven by extrinsic curvature. 

In general, the internal forces appeared in the membrane become
very intense at regions where the divergence of the nematic director increases, 
which correspond to membrane locations where  topological  defects are placed in. Expectedly, textures with zero 
divergence do not induce these forces,  maintaining unchanged the orientation of the nematics over the whole surface.
The bending stress induced by  the nematic director really dominates in curved surfaces by   through of  tangential 
components produced by extrinsic couplings; in fact, if the nematic director is aligned on geodesics, only 
extrinsic effects do contribute.
As regards  twisting stresses, they exclusively arise from extrinsic couplings.
Globally, our results constitute a first achievement on the mechanics of nematic membranes, an intriguing problem 
early captivating  attention of several communities (see reviews in refs. \cite{kamien, giomi}), but still remained 
unresolved because an analytics extremely complex. Although the approach here adopted is not 
mechanical in nature, but  merely geometric indeed,  
the method of auxiliary variables with surface-based constraints  has delivered a complete 
description of the nematic membrane stresses.  \\
\indent Specifically, the forces on circular loops  have been calculated for different textures in 
spherical and a cylindrical settings \cite{chen-kamien}.
For the spherical case, two nematic textures of interest have been analyzed; 
the first  is the case of meridian orientation with finite divergence and a defect  at  each pole, 
and the second  corresponds to parallel orientation with zero divergence.
In both cases, we have obtained the corresponding equilibrium  force equation including the elastic bending 
and the Laplace pressure at mechanical trade-off with the nematic forces. For the meridian
texture, both, splay and  bending modes play a non-trivial role. However,  for the parallel texture, 
only the bending constant intervenes  in the equilibrium equation.
The results here obtained are equivalent to the theory of nematic films in Delaunay surfaces, which has been previously developed by Chen and Kamien \cite{chen-kamien}. Similarly to our results for the "parallel" nematic orientation (Eqs. \eqref{FG}-\eqref{FK} 
for the sphere and Eqs. \eqref{DH}-\eqref{DK} for the cylinder), these authors predicted nematic configurations that are automatically splay-free if lying parallel to the lines of latitude of a surface of revolution \cite{chen-kamien}.  A phase-diagram of stable shapes and topologies was mapped in that theoretical work, a breakthrough that could be generalized using our theory. 
From the equilibrium equation in Eq. \eqref{YYLS}, we can estimate a persistence length scale for 
nematic effects; specifically, $l_\Sigma=2\Sigma R^2/\kappa_1$ in the regime of high membrane tension at high nematic alignment, and $ l_\kappa\approx 2\kappa/\kappa_3$   if bending rigidity governs. 
Taking typical values $\kappa_1\approx 10^{-11}\, N$  and $\kappa_3 \approx 10^{-9}N$, for nanometric 
shells ($R\approx 10\, nm$), we estimate  $l_\kappa \approx 10 \, nm$ in the case of a relatively rigid 
membrane ($\kappa\approx 20  k_B T $), and $l_\Sigma \approx\, 100\, nm$  in the  tensioned case 
membrane ($\Sigma\approx   10^{-2} N/m$). These  predicted scales oversize the systemic dimensions, thus confirming  
the dominance of the molecular director to determine  the configuration of
nanometric-sized nematic shells $(l\gg R)$. Our estimates as agree well  with  previous conclusion by Chen 
and Kamien for  thin film with a revolutionary symmetry,  where anisotropic nematic effects were predicted to 
dominate the membrane shape \cite{chen-kamien}. 
At very high tension, these ordering effects can eventually oversize 
the molecular dimensions up to the macroscopic scale; in the tensioned membrane of a biological cell, for instance 
($R \approx 10\, \mu m; \Sigma\gg  10^{-2} N/m$). Otherwise said,  the characteristic length becomes larger than the 
characteristic cell size $(l_\Sigma\gg R )$, which is due to the strong persistence of the molecular alignment supported by a high lateral tension. On the cylinder,   a texture such that its nematic director follows 
helical trajectories has been analyzed and the forces on parallels and meridians  have been obtained.
Interestingly, there are no induced forces  when the nematic director is aligned along the cylinder axis.
If one considers the director aligned  on parallels, we find that the nematic tension, 
has a finite magnitude  of $\kappa_3/2R $, which contributes to stiff the tube beyond its mere flexural elasticity.  
In the case of a nanometric tube with a nematic membrane ($ R\approx 10\,nm$), for example  a cylindrical vesicle made of oriented polymers, or mitochondrial cristae in the cell biology setting, the membrane tension due to anisotropic ordering should take a value of the order of $\kappa_3/2R\approx 0.1\, N/m$, at least one order of magnitude higher than the typical tensions of isotropic membranes.
Regarding the torque within these nematic membranes, we have found that no
intrinsic torque is induced, in general,  by splaying the nematic director. This is because the splay energy  is purely intrinsic 
to the membrane and  the nematic director is deformed perpendicular to the membrane. As a consequence, splay
contributes only to the external torque. 
Particularly for the sphere, there are no twist contributions too,  so only the bending  plays a role in the torque, 
unlike  the cylinder geometry where  twist and  bend play both a relevant role.

Our geometric theory of the mechanics of nematic membranes builds upon the barest case of fluid isotropic 
membranes, which is harnessed by the Canham-Helfrich theory \cite{stress, deserno, fournier, napoli2010}. 
The early antecedent to a mechanical theory of fluid membranes with in-plane order was focussed on texture topology
\cite{mac}, but not  in membrane curvature, as we elaborate here.  
The {\it ad hoc} introduction of curvature terms in the Frank's energy of nematic shells has been also
considered in studies of structure \cite{vinel}, stability \cite{sonnet}  and geometry-induced distortions in molecular director \cite{vergori}. However, despite the capacity of those approaches in describing the frustrations in nematic ordering elicited by membrane geometry,  a closed theory of the ordering-curvature coupling  completely congruent with geometry is still lacking. The previous work by Chen and Kamien already pointed out the chief role of the nematic bending to elicit spontaneous curvature anisotropy leading to shape instability \cite{chen-kamien}. That study on surfaces of revolution predicted surface buckling, even topological transitions (sphere-torus), at well defined material regimes of tension-to-bending nematic rigidity \cite{chen-kamien}. Unstable scenarios have been also described in previous works as a consequence of either electrically-induced 
molecular tilting \cite{New5} or defect interactions \cite{kardar}. 
Here, we have considered the Frank's energy in a complete geometry-structure coupling schema in which the nematic field is affinelly embedded in the curvature-elasticity field. Our unprecedented achievement represents a major novelty as provides a closed framework to predict the equilibrium distribution of membrane stresses in generalized geometry. This definitively opens a gate to the 
generalized mechanics of nematic membranes,  including the dynamics of shape transformations and stability analyses. \\  
\indent In the biological setting,  the theoretical framework  here developed could contribute a better understanding  of the 
mechanical  remodeling effected by
ordered structures (nematic-like) on flexible membranes. As a relevant example,
our theory could be exploited to model local forces in cellular membranes.
During early stages of cell division, particularly along the cytokinetics processes, specifically aligned actomyosin filaments
present in the division forrow  are known to induce constriction  forces at cell equator. 
This dynamic process arises from a mechanical interplay settled in the membrane between cytokinetic ordering forces and geometric couplings, which finally leads to divisional cell remodeling.
Our FCH-theory of nematic-membranes could pave the unexplored way to link dynamical ordering inside the cell cortex and 
cytokinetic forces.

\section{CONCLUSIONS}\label{CON}
This work addresses a geometric approach to the mechanics of nematic membranes as described by the Frank-Canham-Helfrich (FCH) energy. Using  the method of auxiliary variables, we have given account  for the surface-based constraints that define the geometric characteristics of the coupling connections between membrane curvatures and nematic director. The FCH theory here developed provides an analytic framework to predict the distribution of nematic membrane forces, both tensional stresses and torques, which arise from intrinsic and extrinsic curvature couplings with the molecular director. The nature and the strength of the different couplings have been neatly identified, and their impact in the geometric distortion of the nematic ordering evaluated in simple 
geometrical settings (sphere and cylinder). Although the current theory is essentially geometric, the results approached open a new gate towards the still unavailable mechanical theory of nematic membranes with possible  biological applications. Further work on the theoretical implications of the geometric approach to the FCH field, and its possible extension to a more sophisticated mechanical theory, is ongoing.

\section*{Acknowledgments} JAS thanks to Prof. Francisco Monroy  for hospitality at Universidad Complutense
de Madrid where this work was carried out. The work was supported in part by CONACyT under Becas Sab\'aticas en el Extranjero 
(to JAS), and by Ministerio de Ciencia, Innovaci\'on y Universidades (MICINN, Spain) under grant FIS2015-70339-C2-1-R and by Comunidad de Madrid under grants S2013/MIT-2807, P2018/NMT4389 and Y2018/BIO5207 (to FM).

\appendix
\section{Stress tensor}\label{APX}
In order to obtain the stress ${\bf f}^a$, we  adapt the method of auxiliary variables \cite{auxiliary, jemal-book},  
to implement the geometric constraints related with 
the several  objects  $\Sigma$  describing the surface structure; these are:
$ {\bf e}_a \cdot {\bf n}=0$, $g_{ab}={\bf e}_a \cdot {\bf e}_b$,  ${\bf n}\cdot {\bf n}=1$, ${\bf e}_a = \partial_a {\bf X}$, 
and $K_{ab}={\bf e}_a \cdot \nabla_ b {\bf n}$. The functional to be minimized is written by introducing the corresponding Lagrange 
multipliers as
\begin{eqnarray}
&&H = H_{\rm Frank}  
+   \int dA\,\, 
[ {\bf f}^a \cdot({\bf e}_a -\partial_a {\bf X}) +\frac{\Lambda}{2}( {\bf n}\cdot {\bf n}-1) \nonumber\\
&&+ \Lambda^a ({\bf n}\cdot {\bf e}_a)
+ \Lambda^{ab}(K_{ab}- {\bf e}_a \cdot \partial_a {\bf n}) \nonumber\\
&+& \frac{\lambda^{ab}}{2} (g_{ab}- {\bf e}_a \cdot {\bf e}_b)]. \label{energy}
\end{eqnarray}
In order to applied the auxiliary method, the Frank's energy must be written explicitly in terms of the geometric variables as
\begin{eqnarray}
( {\nabla}\cdot \ets) &=& g^{ab} ( {\bf e}_a \cdot \nabla_b \ets),\nonumber \\
\ets\cdot ({\nabla} \times \ets)  &=& g^{ab} \ets \cdot ( {\bf e}_a \times \nabla_b \ets ),\nonumber \\
(\ets\cdot {\nabla} )  \ets&=& g^{ab} \ets \cdot {\bf e}_a \nabla_b \ets.
\label{FrankSurface}
\end{eqnarray}
If one is interested in closed vesicles then a constraint that fixes the volume enclosed will be needed \cite{jiang}.
The variation  with  respect to the embedding function ${\bf X}$ gives
\begin{equation}
\delta H= \int dA \nabla_a {\bf f}^a \cdot \delta {\bf X} - \int ds\,  l_a {\bf f}^a \cdot \delta{\bf X},
\end{equation}
 so that Euler-Lagrange equation implies the conservation law $\nabla_a {\bf f}^a=0$.  
 On the one hand, variation of the functional $H$
 respect the the tangent vectors gives
 \begin{eqnarray}
 \delta H &=&   \int dA(A^{ac} {\bf e}_c - B^a {\bf n}  ) \cdot \delta {\bf e}_a  \nonumber\\
 &+& \int dA \,   [ {\bf f}^a + \Lambda^a {\bf n} - \Lambda^{ab} \partial_a {\bf n}  - \lambda^{ab} {\bf e}_b ] 
 \cdot \delta{\bf e}_a,
\end{eqnarray}
with coefficients
\begin{eqnarray}
A^{ac}&=&  \kappa_1 \nabla_d \eta^d \nabla^a \eta^c  + \kappa_2  K_\tau  K^a{}_b \eta^b \eta^c_\perp   \nonumber\\
&+& \kappa_3 (\eta^d \nabla_d \eta^b \nabla^a \eta_b 
+ K_\eta \eta^b K_b{}^a )\eta^c \nonumber\\
B^a&=&  \kappa_1 \nabla_d \eta^d  K^{a}{}_c \eta^c  - \kappa_2 K_\tau \eta_{\perp c} \nabla^a \eta^c 
\end{eqnarray}
and $K_\eta= K_{ab}\eta^a\eta^b$, 
$K_\tau= K_{ab}\eta^a_\perp\eta^b$. 
Therefore, we find the stress tensor as
\begin{equation}
{\bf f}^a= ( \lambda^{ac} + \Lambda^{ab} K_b{}^c - A^{ac}) {\bf e}_c - ( \Lambda^a -B^a ) {\bf n}.
\end{equation}
On the other hand, variation  of the functional $H$ with respect to the normal ${\bf n}$ gives
\begin{equation}
\delta H= \int dA [ \lambda \boldsymbol{\eta} + \Lambda{\bf n} + \Lambda^a {\bf e}_a  + \nabla_b( \Lambda^{ab} {\bf e}_b)   )  ] \cdot \delta{\bf n},
\end{equation} 
and because linear independence we have
\begin{eqnarray}
\Lambda - \Lambda^{ab} K_{ab} &=&0, \nonumber\\
\lambda \eta^a + \Lambda^a + \nabla_b \Lambda^{ab}&=&0.
\end{eqnarray}
Here, the  Lagrange multiplier $\Lambda^{ab}$ vanishes because  the extrinsic curvature components $K_{ab}$ do
not appear explicitly  into the energy  $H_{\rm Frank} $.
The fact that the metric $g_{ab}$  appears into $H_{\rm Frank}$,    determines the Lagrange  multiplier 
$\lambda^{ab}= T^{ab}$, where $T^{ab}$ is the tensor defined by the variation  
\begin{equation}
\delta H_{\rm Frank} =-\frac{1}{2} \int dA \, T^{ab}\, \delta g_{ab}.
\end{equation}
In the calculation of this variation, we cast $\delta g^{ab}= -g^{ac}g^{bd} \delta g_{cd}$. After some algebra, we find 
\begin{eqnarray}
T^{ab}_S &=& \kappa_1[ ({\bf e}^c\cdot \nabla_c \boldsymbol{\eta})( {\bf e}^a \cdot \nabla^b {\bf \boldsymbol{\eta}}  +  {\bf e}^b \cdot \nabla^a {\bf \boldsymbol{\eta}}   )  
 - \frac{g^{ab}}{2}(  {\bf e}^c\cdot \nabla_c \boldsymbol{\eta} )^2 ], \nonumber\\ 
&=&\kappa_1 (\nabla_c\eta^c) \Big[  (\nabla^a\eta^b +\nabla^b\eta^a ) -\frac{ g^{ab}}{2}(\nabla_d\eta^d) 
\Big] , \nonumber\\
T^{ab}_W&=& \kappa_2 [  K_\tau \boldsymbol{\eta}\cdot ( {\bf e}^a\times \nabla^b \boldsymbol{\eta} +   {\bf e}^b\times \nabla^a \boldsymbol{\eta} )  \nonumber\\
&&- \frac{g^{ab}}{2}  ( g^{cd} \boldsymbol{\eta}\cdot ( {\bf e}_c \times \nabla_d \boldsymbol{\eta} ) )^2   ], \nonumber\\
&=& \kappa_2[ K_\tau ( K_{cb}\eta^c \eta_{\perp a} +  K_{ca}\eta^c \eta_{\perp b}  )  
- \frac{g^{ab}}{2}  K_\tau^2   ], \nonumber\\
T^{ab}_B&=& \kappa_3 \,  \eta^d \nabla_d \boldsymbol{\eta}\cdot   [ \eta^a \nabla^b  \boldsymbol{\eta} 
+ \eta^b \nabla^a  \boldsymbol{\eta} - 
\frac{g^{ab}}{2} \eta^c \nabla_c \boldsymbol{\eta} ] \nonumber\\
&=&\kappa_3\, [ K_\eta\eta^c ( \eta^a K^b{}_c  + \eta^bK^a{}_c ) \nonumber\\
&+& \eta^d \nabla_d \eta^c( \eta^a\nabla^b \eta_c + \eta^b \nabla^a\eta_c )\nonumber\\
&-& \frac{g^{ab}}{2}( \eta^d \eta^e \nabla_d \eta^c \nabla_e \eta_c + K_\eta^2   )  ].
\end{eqnarray}
The stress is then written as
\begin{equation}
{\bf f}^a = (T^{ac}  -A^{ac}) {\bf e}_c -(\Lambda^a - B^a ) {\bf n},
\end{equation}
where the multiplier $\Lambda^a$ will be determined once the multiplier  $\lambda$ does. To find $\Lambda^a$
we calculate the variation of the energy Eq. \eqref{energy} under deformations of the field itself; up to a
boundary terms, we have
\begin{equation}
\delta H= \int dA (C^{a} {\bf e}_a + D{\bf n} ) \cdot \delta\boldsymbol{\eta} + \int dA (\beta \boldsymbol{\eta}  + \lambda {\bf n} )  \cdot \delta \boldsymbol{\eta} 
\end{equation}
where 
\begin{eqnarray}
C^{a}&=& - {\kappa}_1  g^{ab}\nabla_b\nabla_c  \eta^c  +  
{\kappa}_2 K_\tau ( 2 K_{bc} \eta^b \varepsilon^{ca}       +    K \eta^a_\perp  ) \nonumber\\
&&+{\kappa}_3  [ \eta^b \nabla_b\eta_d (\nabla^a\eta^d )    - \nabla_b \eta^b (\eta^c \nabla_c \eta^a  ) - \eta^b \nabla_b \eta^c (\nabla_c\eta^a ) \nonumber\\
&&- \eta^b\eta^c \nabla_b\nabla_c\eta^a + 2 K_\eta K^a{}_b \eta^b ]\nonumber\\
D&=& {\kappa}_1  \nabla_a \eta^a K + {\kappa}_2 ( 2K_\tau  \nabla_a \eta^a_\perp + \eta^a_\perp\nabla_a K_\tau ) \nonumber\\
&&+ \kappa_3 [  \nabla_b \eta^b K_\eta + \eta^b \nabla_b \eta^a K_{ad} \eta^d  \nonumber\\
&&+  \eta^b \eta^c ( \nabla_c \eta^d K_{bd} + \nabla_b K_{ cd}\eta^d + K_{cd} \nabla_b \eta^d      )        ]
\end{eqnarray}
with the  missing Lagrange multipliers being $\lambda= -D$ and $\beta= -C^a \eta_a$,  so that $\Lambda^a = D \eta^a $. 
Then, we can write
\begin{equation}
{\bf f}^a = (T^{ac}  -A^{ac}) {\bf e}_c -(D \eta^a - B^a ) {\bf n},
\end{equation}
where the values of $T^{ab}$, $A^{ab}$ and $B^a$ should be substituted  by their corresponding expressions above expanded.
Thus, we find that
the splay stress tensor becomes proportional to the divergence of the director field. It can  be 
written as
\begin{eqnarray}
{\bf f}^a_S&=&\kappa_1 \nabla_d \eta^d \left( \nabla^c \eta^a  -
 \frac{g^{ac}}{2}\nabla_d\eta^d \right) {\bf e}_c\nonumber\\
&-&\kappa_1\nabla_c \eta^c (K    \eta^a -  K^a{}_c \eta^c) {\bf n},
\end{eqnarray}
where the tangential components have intrinsic information of the surface through the metric $g_{ab}$ and the
covariant derivatives of $\eta^a$. The normal projection has, instead, coupling with  extrinsic curvature.
The splay force per unit length can then be written as
\begin{equation}
{\bf f}^a_S l_a= F_T {\bf T}+ F_l {\bf l}+ F_n {\bf n}, 
\end{equation}
where the projections are given by
\begin{eqnarray}
F_T&=& \kappa_1 \nabla_d\eta^d T^c l^a \nabla_c \eta_a, \nonumber\\
F_l&=& \kappa_1 \nabla_d\eta^d \left( l^al^c  \nabla_c\eta_a - \frac{1}{2} \nabla_b\eta^b \right),\nonumber\\
F_n&=& -\kappa_1\nabla_d \eta^d (  K \eta_a l^a - K_{ab}l^a\eta^b ).
\end{eqnarray}
As expected, the extrinsic coupling appears only in the normal projection. 
The twist stress tensor is identified as:
\begin{eqnarray}
{\bf f}_W^a &=&\kappa_2 K_\tau (K^{cb} \eta_c \eta_\perp^a 
- \frac{g^{ab}}{2}K_\tau  ){\bf e}_b \nonumber\\
&-& \kappa_2[(2K_\tau \nabla_c\eta^c_\perp + \eta^c_\perp \nabla_c K_\tau  ) \eta^a \nonumber\\
&+& K_\tau \eta_{\perp c} \nabla^a \eta^c ] {\bf n},
\end{eqnarray}
where all the terms have coupling with extrinsic curvature.
In this case,  we have
\begin{eqnarray}
F_T&=& \kappa_2 K_\tau K_{cb}\eta^c T^b \eta_\perp^a l_a, \nonumber\\
F_l&=& \kappa_2 K_\tau \left(  K_{cb} \eta^c l^b  \eta_\perp^a l_a - \frac{K_\tau}{2}   \right), \nonumber\\
F_n&=&- \kappa_2[(2K_\tau \nabla_c\eta^c_\perp + \eta^c_\perp \nabla_c K_\tau  ) \eta^a l_a \nonumber\\
&+& K_\tau \eta_{\perp c} l^a\nabla_a \eta^c ].
\end{eqnarray}
Finally, the bend stress tensor can be obtained as
\begin{eqnarray}
{\bf f}_B^a &=& \kappa_3 \Big[K_\eta \eta^c \eta^a K^b{}_c + \eta^a\eta^d  \nabla_d \eta^c\nabla^b \eta_c \nonumber\\
&-& \frac{g^{ab}}{2} (\eta^d\eta^e \nabla_d\eta^c \nabla_e \eta_c + K_\eta^2)  \Big] {\bf e}_b \nonumber\\
& -& \kappa_3 \Big[  \nabla_b \eta^b K_\eta + 3\eta^b \nabla_b \eta^c K_{cd} \eta^d  \nonumber\\
&+&  \eta^b \eta^c\eta^d  \nabla_b K_{ cd}  \Big] \eta^a {\bf n}.
\end{eqnarray}
Therefore, we can obtain the projections as
\begin{eqnarray}
F_T&=& \kappa_3[ K_\eta K_{bc}T^b \eta^c + \eta^d \nabla_d \eta^c T^b\nabla_b \eta_c   ] \eta^a l_a \nonumber\\
F_l&=& \kappa_3[ K_\eta K_{bc}l^b \eta^c + \eta^d \nabla_d \eta^c l^b\nabla_b \eta_c   ] \eta^a l_a \nonumber\\
&-& \frac{\kappa_3}{2}\left( \eta^d\eta^e \nabla_d\eta^c \nabla_e \eta_c + K_\eta^2 \right) \nonumber\\
F_n&=&- \kappa_3 \Big[  \nabla_b \eta^b K_\eta + 3\eta^b \nabla_b \eta^c K_{cd} \eta^d  \nonumber\\
&+&  \eta^b \eta^c\eta^d  \nabla_b K_{ cd}  \Big] \eta^a l_a
\end{eqnarray}

\bibliographystyle{abbrv}
\bibliography{refs}

\newpage

\end{document}